\documentclass[a4paper,12pt]{elsarticle}
\usepackage{graphicx}
\DeclareGraphicsExtensions{.eps.png.pdf}
\usepackage{indentfirst}
\usepackage{amsthm}
\usepackage{rotating}
\begin{document}
\begin{frontmatter}
\author{Demetris P.K. Ghikas\corref{corr}}
\ead{ghikas@physics.upatras.gr}
\cortext[corr]{Corresponding author}
\author{Fotios D. Oikonomou}
\address{Department of Physics University of Patras,\\ Patras 26500, Greece}
\title{Towards an Information Geometric characterization/classification of Complex Systems. \\
I. Use of Generalized Entropies}
\begin{abstract}
Using the generalized entropies which depend on two parameters we propose a set of quantitative characteristics derived from the Information Geometry based on these entropies. Our aim, at this stage, is modest, as we are first constructing some fundamental geometric objects. We first establish the existence of a two-parameter family of probability distributions. Then using this family we derive the associated metric and we state a generalized Cramer-Rao inequality. This gives a first two-parameter  classification of complex systems. Finally computing the scalar curvature of the information manifold   we obtain a further discrimination of the corresponding classes.
Our analysis is based on the two-parameter family of generalized entropies of Hanel and Thurner (2011).
\end{abstract}
\end{frontmatter}
\newpage
\section{Introduction}
\indent
\subsection*{The problem of characterization and classification of complex systems}
Complex Systems are ubiquitous in nature and in man made systems. They are objects of study in natural sciences, in social and economic models and in mathematical and  information constructions. But despite this extensive activity there is still lacking a universal consensus of the meaning of  the word "complex". And though it is understood that complex is different from complicated there is no generally accepted definition of "complex systems" let alone quantitative characterization and qualitative classification.  There are many "system  approaches" and few "axiomatic approaches".  In the first, in the framework of a particular discipline and concrete examples there is an abstraction which encapsulates some common and  definite properties. Typical examples are the Theory of Dynamical Systems, Neuroscience and Financial Markets. This is a case study approach which cannot cover all systems qualified as complex \cite{nicol1,rotten,monc}. The axiomatic approaches, certainly emanating from case studies, go further to identify universal properties \cite{zad,lady,mag}. In the system approaches there exist some quantitative measures, computable or mainly non-computable, but cover a particular area.  In the abstract approaches there are qualitative characterizations but no algorithmic definition. But there is a common concept that plays a fundamental role in this activity. This is the stochasticity and the statistical behavior. Since it is not the purpose of this work to review and comment on definitions of complexity we focus on the possibility to use statistical tools quantitatively. A well established framework is based on the concept of entropy and its various generalizations. 
There is an extensive literature on the use of entropy in connection to case studies of complex systems \cite{koore,mere,gaud,tsall1,tsall2,tsall3,beck,oik,late}. But there is a particular generalization of entropy that it is assumed to classify complex systems, the so called (c,d)- entropies of Hanel and Thurner \cite{hanel1}. Our approach is the use of this entropy to construct particular Information Manifolds. From these we construct geometric quantities with which we classify complex systems. There have been similar constructions of Information Geometry, but they are based on single parameter generalizations of entropy 
\cite{amari1,amari2,naud1,naud2,wada,mats,hars,masi,penn,abe,ay,ber}. 
  
\indent
\subsection*{Generalized Entropies}
\indent
The development of Statistical Mechanics and its associated thermodynamic limit was based on the thermodynamic behavior of physical systems with short range forces. The fundamental tool for the theoretical analysis is the Boltzman-Gibbs Entropy. Inherent in this formalism is the Legendre structure which incorporates the duality between extensive and intensive thermodynamic quantities. Shannon and later Khinchin were the first to deal with entropy in a rigorous way. This is based on four axioms which uniquely determine the well known functional form of the  entropy. Key concepts are the extensivity and additivity properties which for the Boltzman-Gibbs entropy coincide. After Renyi's non standard entropy functional, Tsallis \cite{tsall2,tsall3} proposed a one-parameter entropy functional which it is more suitable for systems with long range forces. This entropy does not satisfy the property of additivity. After that there has been a host of different entropic functionals, constructed under particular assumptions and satisfying certain conditions supposed to hold for particular systems. We are interested on the two-parameter entropy functional of Hanel and Thurner \cite{hanel1} because it is proposed as a mathematical tool for the classification of complex systems. Now for a given entropy functional one can obtain the probability distribution function that minimizes this functional under the Maximum Entropy Principle. But in any variational procedure one needs the escorting conditions that enter with their Lagrange multipliers. All these entropies produce their associated probability distributions, uniform, exponential and so on. In this part of our work we use the distribution associated with the two-parameter entropy of  Hanel and Thurner \cite{hanel1,hanel2,hanel3} to construct our information geometric quantities.

\subsection*{Information Geometric tools}
Information Geometry emerged as a practical geometric framework in the theory of parameter estimation in mathematical statistics \cite{amari1}. For a given statistical model, that is a given class of probability measures there is associated an information manifold and a Riemannian metric. This metric enters in the estimation procedure through the Cramer-Rao inequality giving the possible accuracy of an estimator of the parameters of the model. Further on, one may define non-Riemannian connections which offer a deeper analysis of the estimation procedure. Our work is based on the geometric quantities emerging in the Information Geometry which is based on the two-parameter entropy functional of Hanel and Thurner \cite{hanel1}. In this paper we construct the information manifold, we prove the  appropriate properties of the distribution function and use the Cramer-Rao Inequality and the scalar curvature to construct certain plots that differentiate between various classes of complex systems. In a subsequent paper we make a similar classification using different objects of the information geometry. This classification seems to be more discriminating but fails for certain parameter values, a problem that seems not to be of a mere technical origin.   
\\
\indent
In Paragraph 2 we introduce in a minimal way the necessary definitions and geometric quantities of Information Geometry. We present the Cramer-Rao Inequality which we use for our classification, as well as the connection used to compute the scalar curvature that quantifies our classes. In Paragraph 3 we present the main forms of the proposed generalized entropies in the literature with a short discussion and comments on their nature and applicability. Then the generalized entropy of Hanel and Thurner \cite{hanel1} is introduced with few comments on its derivation and properties. In Paragraph 4 we present our results. First we state some theorems which prove the appropriateness of the generalized distribution function. Then we compute the Riemannian metric for the (c,d)-entropy and present the dependence of the Cramer-Rao bound on the values of c and d. Our graphs indicate the differences between various classes. Finally we compute the scalar curvature which clearly indicates that various classes of complex systems have differences in their information manifolds. In the last Paragraph we discuss our approach and comment on its applicability and possible extension. In the Appendix some extra formulas are given and the proofs of the theorems.

\section{Basic concepts of Information Geometry}
\subsection{Geometry from probability distributions and the Cramer-Rao Inequality}
\indent 
Here we present only the necessary concepts in order to establish the notation. We refer to the bibliography for the details \cite{amari1,amari2}. Let
\begin{equation}
S=\{p_{\xi}=p(x;\xi) | \xi = [\xi^{1},...,\xi^{n}]\in \Xi\}
\end{equation}
be a parametric family of probability distributions on $\mathcal{X}$. This is an n\-dimensional parametric statistical model. Given the N observations $x_{1},...,x_{N}$ the Classical Estimation Problem concerns the statistical methods that may be used to detect the true distribution, that is to estimate the parameters $\xi$. To this purpose, an appropriate estimator is used for each parameter. These estimators are maps from the parameter space to the space of the random variables of the model. The quality of the estimation is measured by the variance -covariance matrix $V_{\hat{\xi}}=[v_{\xi}^{ij}]$ where
\begin{equation}
v_{\xi}^{ij}= E_{\xi}[(\hat{\xi}(X)-\xi^{i})(\hat{\xi}(X)-\xi^{j})]
\end{equation} 
Suppose that the estimators are unbiased, namely
\begin{equation}
E_{\xi}[\hat{\xi}(X)]=\xi , \quad \forall \xi \in \Xi
\end{equation}
Then a  lower bound for the estimation error is given by the Cramer-Rao inequality
\begin{equation}
V_{\xi}(\hat{\xi})\geq G(\xi)^{-1}
\end{equation}
where $G(\xi)=[g_{ij}(\xi)]$ 
\begin{equation}
g_{ij}(\xi)=E_{\xi}[\partial_{i}l(x;\xi)\partial_{j}l(x;\xi)]
\end{equation}
the Classical Fisher Matrix with 
\begin{equation}
l_{\xi}=l(x;\xi)=lnp(x;\xi)
\end{equation}
the score function.
As it has been shown the Fisher Matrix provides a metric on the manifold of classical probability distributions. This metric, according to the theorem of Cencov \cite{cencov}, is the unique metric which is monotone under the transformations of the statistical model. This means that if the map $F : \mathcal{X} \to \mathcal{Y}$ induces a model $S_{F}=\{q(y;\xi)\}$ on $\mathcal{Y}$ then 
\begin{equation}
G_{F}(\xi)\leq G(\xi)
\end{equation} 
That is, the distance of the transformed distributions is smaller than the original distributions. Thus monotonicity of the metric is intuitively related to the fact that in general we loose distinguishability of the distributions from any transformation of the information.
\newline
\indent
The metric defined in this way is the ordinary Fisher metric. Using the Levi-Civita connection the corresponding Riemannian structure is  constructed. In this geometry the scalar curvature is a quantification of the information manifolds. But there is a further development connected with the existence of connections different from Levi-Civita. These are certain pairs of connections satisfying a duality property with respect to the Fisher metric and playing a fundamental role in the estimation theory. An important case is the dually flat connections. 
\newline
\indent
\subsection{Geometry from Divergencies}
A further extension is the derivation of the differential structure from relative entropies or divergence functions. These are quasi-distances and particular cases have been used with various names. Let p,q be distribution functions considered as points in an information manifolds. A divergence $D(p\|q)$ satisfies the property
\begin{equation}
D(p\|q) \geq 0 \quad and \quad D(p\|q) = 0 \quad iff \quad p=q
\end{equation}   
Now considering the function $D(p\|p+dp)$ and expanding to third order we get a metric and a connection characterized by D :

\begin{equation}
g_{ij}^{D} = - \partial_{i}\partial_{j}^{'}D(p\|p^{'})|_{p^{'}=p}
\end{equation}

\begin{equation}
\Gamma_{ij,k}^{D} = - \partial_{i}\partial_{j}\partial_{k}^{'}D(p\|p^{'})|_{p^{'}=p}
\end{equation}
where $\partial_{i} = \frac{\partial}{\partial\xi^{i}}$ and $\partial_{i}^{'} = \frac{\partial}{\partial\xi^{'i}}$. A fundamental concept of great practical usefulness in the estimation theory is the duality. Given a metric and two connections $(g,\nabla, \nabla^{*})$  the connections are dual with respect to the metric if 
\begin{equation}
\partial_{k}g_{ij} = \Gamma_{ki,j} + \Gamma_{kj,i}^{*}
\end{equation} 
holds. From the geometry coming from a divergence a dual structure is obtained by defining $\Gamma_{ij,k}^{D^{*}} = - \partial_{k}\partial_{i}^{'}\partial_{j}^{'}D(p||p^{'})|_{p^{'}=p}$.
There is a general family of divergences, the so called f-divergences, which are generalizations of the known Kullback - Leibler divergence. In the statistical applications a special role is played  by the dually flat connections. In this case there exist dual coordinate systems on the manifold, $[\theta^{i}]$ , $[\eta_{j}]$ and  functions $\psi$  and $\phi$ such that
\begin{equation}
\theta^{i} = \partial^{i}\phi \quad,\quad \eta_{i} = \partial_{i}\psi \quad, \quad g_{ij} = \partial_{i}\partial_{j}\psi \quad, \quad g^{ij} = \partial^{i}\partial^{j}\phi
\end{equation}
This is a Legendre Transformation with the corresponding potential function $\psi$ and $\phi$. \\
\indent
There is a canonical divergence which is uniquely defined for dually flat manifolds 

\begin{equation}
D(p||q) \equiv \psi(p) + \phi(q) -\theta^{i}(p)\eta_{i}(q)
\end{equation}
Exponential families have an inherent dually flat structure. And this offers a natural root to construct geometries for generalized exponentials which are related to generalized entropies.
\cite{amari1}
\section{Generalized Entropies and Complex Systems}
\indent
\subsection{Generalized Entropies}
\indent
Assuming the four Shannon-Khimchin axioms it is proved that there exists a unique entropy functional, the Boltzmann-Gibbs Entropy

\begin{equation}
S[p] = -\sum_{j\in J}p(j)lnp(j) \quad , \quad  \sum_{j\in J}p(j) = 1
\end{equation}
These axioms are plausible assumptions abstracted from the typical behavior of thermodynamic systems and the role of thermodynamic entropy. But after the statistical foundation of thermodynamics and the association of entropy with information theory, it became necessary to look for other functionals which were thought to cover more general systems than the simple ones like perfect gases, and more generally systems with long range interactions. And though it is expected that in the thermodynamic limit to have functionals with a universal form, it is evident that for small systems one needs functionals dependent on parameters. These parameters, not having always a transparent connection with the empirical properties of the systems, nevertheless, offered a minimal parametric generalization of the Boltzmann-Gibbs functional as an information theoretic tool. One of the earliest generalizations is the Renyi's Entropy

\begin{equation}
S^{q}[p] = \frac{1}{1-q}ln ( \sum_{j}p(j)^{q} )
\end{equation}   
Later on Tsallis\cite{tsall1,tsall2,tsall3}, in relation to the theory and practice of fractals introduced his entropy

\begin{equation}
S_{q}^{Tsallis}[p] = \frac{1}{1-q}( \sum_{j}p(j)^{q}-1 )
\end{equation}
a form that had been introduced earlier for mathematical reasons. There after a host of other forms of entropic functionals were introduced associated with particular properties of complex statistical systems. All these entropies, assuming a form of Maximal Entropy Principle give rise to probability distributions which depend on the parameter of entropy. In general these are generalized exponentials which are the inverse functions of generalized logarithms. These generalized exponentials, assumed to be particular exponentials of probability distributions may be used to construct information geometric objects. In this work we use the two-parameter entropic functional of Hanel and Thurner to construct our geometric tools. 
 
\subsection{A two-parameter Generalized Entropy and Complex Systems}
\indent

Given the fact that the four Shannon-Khinchin Axioms impose a unique form for the entropy, which is the Boltzmann-Gibbs functional, Hanel and Thurner, seeking a generalization to the case of a functional not satisfying additivity had to abandon the relevant axiom. Their analysis produced a two-parameter entropic functional of the form

\begin{equation}
S_{c,d}[p] =  \frac{e\sum_{i}^{W}\Gamma(d+1,1-clnp_{i})}{1-c+cd} - \frac{c}{1-c+cd}
\end{equation}
where W is the number of potential outcomes and $\Gamma(a,b) = \int_{b}^{\infty}dt t^{a-1}exp(-t)$ the incomplete Gamma-function.
 The Bolzmann-Gibbs entropy is recovered for (c,d) = (1,1), while for the Tsallis entropy we have (c,d) = (c,0).
\newline
\indent
The maximizing distribution function is the generalized exponential
\begin{equation}
\mathcal{E}_{c,d,r}(x) = e^{-\frac{d}{1-c}[W_{k}(B(1-x/r)^\frac{1}{d}) - W_{k}(B)]} 
\end{equation}
where $r = (1-c+cd)^{-1}$ and $B = \frac{(1-c)r}{1-(1-c)r}exp(\frac{(1-c)r}{1-(1-c)r})$. The function $W_{k}$ is the k-th branch of the Lambert W function which is a solution of the equation $x = W(x)exp(W(x))$. This generalized exponential is the inverse function of the generalized logarithm (under appropriate conditions)

\begin{equation}
\Lambda_{c,d,r}(x) = r-rx^{c-1}[1-\frac{1-(1-c)r}{rd}ln x ]^{d}
\end{equation}

\section{Results}

\subsection{The (c,d)-exponential family}
\indent
Amari and Ohara \cite{amari2} studied the geometry of q-exponential family of probability distributions. We repeat this analysis for the (c,d)-exponential family. 
First it is easily seeing that for $x \in (0,1]$ and c,d,r  real if $\frac{1-(1-c)r}{dr}\geq 0$ then the generalized logarithm $\Lambda_{c,d,r}(x)$ is a real function.
This is connected to the conditions given by Hanel and Thurner
\begin{eqnarray}
d>0 &:& r<\frac{1}{1-c} \\
d=0 &:& r=\frac{1}{1-c} \\
d<0 &:& r>\frac{1}{1-c}
\end{eqnarray}
Now the distribution $\mathcal{E}_{c,d,r}(x)$ is characterized as exponential family if \\$p(x,\theta) = \mathcal{E}_{c,d,r}(x_{i}\theta^{i} - \psi(\theta))$ or equivalently $\Lambda_{c,d,r}(p(x,\theta)) = x_{i}\theta^{i} - \psi(\theta)$. That this distribution is exponential is proved in our first theorem
\vspace{12pt}
\\
\textbf{Theorem 1}
\newline
\indent
\textit{The family with the discrete distribution $p = (p_{0},p_{1},...,p_{n})$ with \\
$p_{i}= Prob(x=x_{i})$ and $p_{0} =1- \sum_{i=1}^{n} p_{i}$ has the structure of a (c,d) exponential family with}
\begin{equation}
\theta^{i} = rp_{0}^{c-1}\Biggl[1-\frac{1-(1-c)r}{dr}ln p_{0}\Biggr]^{d} - rp_{i}^{c-1}\Biggl[1-\frac{1-(1-c)r}{dr}ln p_{i}\Biggr]^{d} 
\end{equation}

\begin{equation}
\mathit{x_{i} = \delta_{i}(x) = \left\{
  \begin{array}{lr}
	1 &x = x_{i}\\
	0 &x \neq  x_{i}
	\end{array}
	\right.
	i = 1,...,n}
\end{equation}

\begin{equation}
\mathit{\psi(\theta) = -\Lambda_{c,d,r}(p_{0})}
\end{equation}
\newline
Let the function
\begin{equation}
\Delta(x) = \frac{1}{r(1-c)}\mathcal{E}_{c,d,r}(x)\frac{W(B(1-x/r)^{1/d})}{1+W(B(1-x/r)^{1/d})}(1-\frac{x}{r})^{-1}
\end{equation}
We get 
\begin{equation}
\frac{\partial\psi}{\partial\theta^{i}}\equiv \partial_{i}\psi
= \frac{\int x\Delta(x_{j}\theta^{j}-\psi(\theta))dx}{\Delta(x_{j}\theta^{j}-\psi(\theta))dx}
\end{equation}

From this we have
\vspace{12pt}
\\
\textbf{Theorem 2}
\newline
\indent
\textit{The function $\psi(\theta)$ is convex for the values of c,d,r for which}
\begin{equation}
\mathit{\Delta(x_{j}\theta^{j}-\psi(\theta)) \geq 0 \quad , \quad \Delta^{'}(x_{j}\theta^{j}-\psi(\theta)) \geq 0 }
\end{equation}
\subsection{The (c,d)-information metric}
\indent
We define the functions
\begin{equation}
K(x) = x^\frac{c-1}{d}\left(1-\frac{1-(1-c)r}{rd}lnx\right)      
\end{equation}
so that

\begin{equation}
\Lambda_{c,d,r}(x) = r - rK^{d}(x)
\end{equation}

and
\begin{equation}
h(p) =  \int\Delta(\Lambda(p(x,\theta)))dx \equiv \int\Delta(x_{j}\theta^{j}-\psi(\theta))dx
\end{equation}
For a discrete distribution we have

\begin{equation}
h(p) = \sum_{i=0}^{n}p_{i}\frac{1}{r(1-c)}\frac{W(BK(p_{i}))}{1+W(BK(p_{i})} K^{-d}(p_{i})
\end{equation}
We define the (c,d)-divergence as a canonical divergence

\begin{equation}
D_{c,d,r}(p(x,\theta_{1}),p(x,\theta_{2}) = \psi(\theta_{2}) -  \psi(\theta_{1}) - [\partial_{i}\psi(\theta_{1})](\theta_{2}^{i}- \theta_{1}^{i})             )
\end{equation}

Then we have
\vspace{12pt}
\\
\textbf{Theorem 3}\\
\indent
\textit{For two discrete distributions $p = (p_{0},p_{1},...,p_{n})$ , $q = (q_{0},q_{1},...,q_{n})$ we have for the (c,d)-divergence the expression}
\begin{equation}
\mathit{D_{c,d,r}(p,q) = \frac{1}{(1-c)h(p)}\sum_{i=0}^{n}\frac{p_{i}W(BK(p_{i}))}{1+W(BK(p_{i}))}(K^{-d}(p_{i})K^{d}(q_{i})-1)}
\end{equation}
\newline
Finally defining the metric for a discrete distribution
\begin{equation}
g_{ij}(p) = \frac{\partial^{2}}{\partial q_{i}\partial q_{j}}D_{c,d,r}(p,q)|_{q=p}
\end{equation}
we have
\vspace{12pt}
\\
\textbf{Theorem 4}
\begin{equation}
\mathit{g_{ij}(p) = \frac{1}{(1-c)h(p)}(H(p_{0})+\delta_{ij}H(p_{j}))}
\end{equation}
\textit{where}
\begin{equation}
\mathit{H(x) = x\frac{W(BK(x))}{1+W(BK(x))}(d(d-1)K^{-2}(x)[K^{'}(x)]^{2} + dK^{-1}(x)K^{''}(x))}
\end{equation}
\newline
It can be seen that this metric for d=1 and $c \to 1$ gives the Fisher metric.

\subsection{Cramer-Rao Inequalities for Complex Systems}
Here we follow the analysis of Nauds\cite{naud1,naud2}. 
A new information metric can be defined using two distributions $P_{\theta}=P_{\theta}(x,\theta)$ and $p_{\theta} = p_{\theta}(x,\theta)$

\begin{equation}
\tilde{g}_{ij}(\theta) = \int_{\omega}d\mu(x)\frac{1}{P_{\theta}(x)}\frac{\partial p_{\theta}}{\partial \theta^{i}}\frac{\partial p_{\theta}}{\partial \theta^{j}}
\end{equation}
For the discrete distribution we get

\begin{equation}
\tilde{g}_{ij}(\theta) = \frac{1}{P_{0}} + \delta_{i,j}\frac{1}{P_{i}}
\end{equation}
The following theorem holds
\vspace{12pt}
\\
\textbf{Theorem} (Nauds\cite{naud1,naud2})
\newline
\indent
\textit{Let two families of probability distributions $P_{\theta}=P_{\theta}(x,\theta)$ and $p_{\theta} = p_{\theta}(x,\theta)$ and the corresponding expectations $F_{\theta}$ and $E_{\theta}$. Let c be an estimator of $p_{\theta} = p_{\theta}(x,\theta)$ with scale function F, that is $E_{\theta}[c_{k}]=\frac{\partial}{\partial\theta^{k}}F(\theta)$. Assume the regularity condition}
\begin{equation}
\mathit{F_{\theta}\left[\frac{1}{P_{\theta}(x)}\frac{\partial}{\partial\theta^{k}}p_{\theta}\right] = 0}
\end{equation}
\textit{holds. Let the metric $\tilde{g}_{ij}(\theta)$ introduced above. Then for all u,v in $\rm{R}^{n}$}
\begin{equation}
\frac{u^{k}u^{l}\left[F_{\theta}[c_{k}c_{l}]-F_{\theta}[c_{k}]F_{\theta}[c_{l}]\right]}{\left[u^{k}v^{l}\frac{\partial^{2}}{\partial\theta^{k}\partial\theta^{l}}F(\theta)\right]^{2}}\geq \frac{1}{v^{k}v^{l}\tilde{g}_{ij}(\theta)}
\end{equation}
\newline
We apply this theorem, which gives a generalization of the classical Cramer-Rao inequality using the (c,d)-information metric with the associated escort distribution. For the latter we choose
\begin{equation}
P_{i} = \frac{1}{N(p)}\frac{(1-c)h(p)}{H(p_{i})}
\end{equation} 
where 
\begin{equation}
N(p) = \sum_{i=0}^{n}\frac{(1-c)h(p)}{H(p_{i})}
\end{equation}
the normalization factor.
With this choice the theorem gives
\begin{equation}
\frac{u^{k}u^{l}\left[F_{\theta}[c_{k}c_{l}]-F_{\theta}[c_{k}]F_{\theta}[c_{l}]\right]}{\left[u^{k}v^{l}\frac{\partial^{2}}{\partial\theta^{k}\partial\theta^{l}}F(\theta)\right]^{2}}\geq \frac{1}{N(p)v^{k}v^{l}g_{ij}(\theta)}
\end{equation}

Now for a first indication of the dependence of the generalized Cramer-Rao bound on the parameters of various complexity classes (a la Hanel and Thurner), we compare this bound with the corresponding classical one given by the Fisher metric, for the simplest case of n=1. Then we have an one dimensional manifold with coordinate the probability $p_{1}$. Of course $p_{0} = 1- p_{1}$. Then we have

\begin{equation}
f(c,d,p_{1}) \equiv \frac{1}{N(p)g(p)} = \frac{1}{\left(\frac{1}{H(p_{0})} + \frac{1}{H(p_{1})}\right)\left(H(p_{0}) +  H(p_{1})\right)}
\end{equation} 
The corresponding function for the Fisher metric is

\begin{equation}
f_{Fisher}(p_{1}) = \frac{1}{g_{Fisher}} = \frac{1}{\frac{1}{p_{0}} + \frac{1}{p_{1}}}
\end{equation}
We make the Hanel-Thurner choice for r namely 
\begin{equation}
r = \frac{1}{1-c+cd}
\end{equation}

Our elementary first "visual" classification is made with 2d and 3d graphs depending on $c,d,p_{1}$. The "classifying quantity" is the difference between the (c,d) bound and the Fisher bound. But some classes have fixed values of one or both of c,d and others are connected with intervals of values. For this reason we plot the function
\begin{equation}
I_{mean} = mean_{p_{1},c,d}\left\{f(p_{1}) - f_{Fisher}(p_{1})\right\}
\end{equation}
where "mean" means average over the corresponding non-constant parameters c,d. In the figure captions it is explained over  which parameter is the average. In the figures we show how the Cramer-Rao bound differs from the one given by the Fisher metric for various classes of complex systems. 

\begin{figure}[htp]
	\centering
\scalebox{0.7}{\includegraphics[angle=0]{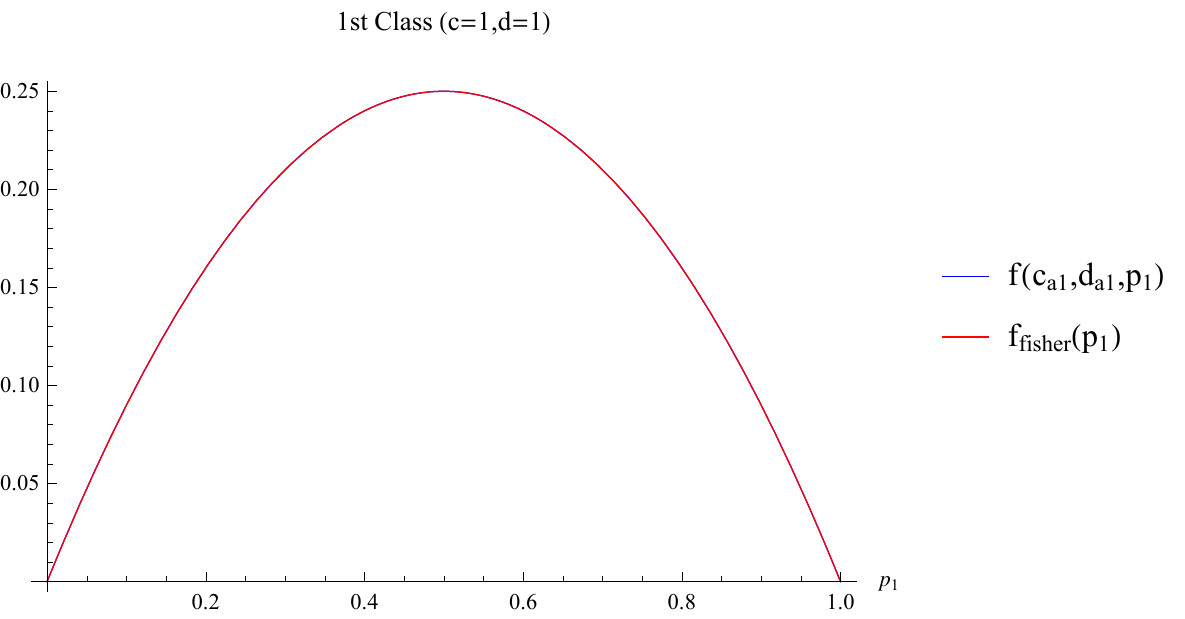}}
\caption{1st class : c =1 , d = 1}
	\label{fig:2D class1}
\end{figure}

\begin{figure}[htp]
	\centering
\scalebox{0.7}{\includegraphics[angle=0]{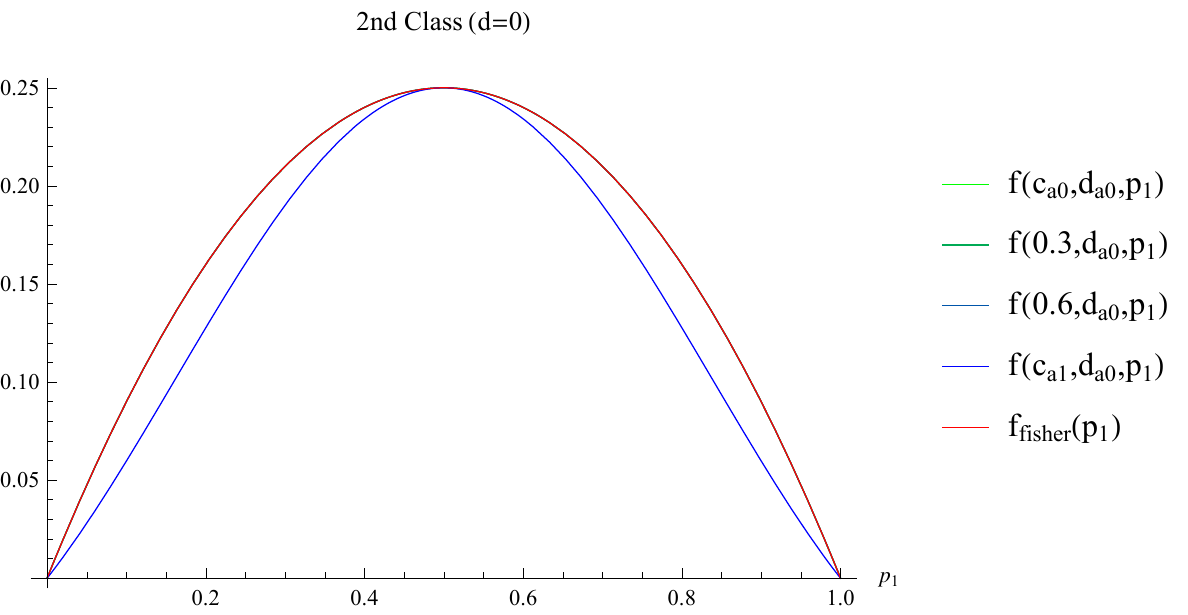}}
\caption{2nd class : (c,d) = (0,0) , (0.3,0) , (0.6,0) , (1,0) }
	\label{fig:2D class 2}
\end{figure}

\begin{figure}[htp]
	\centering
\scalebox{0.7}{\includegraphics[angle=0]{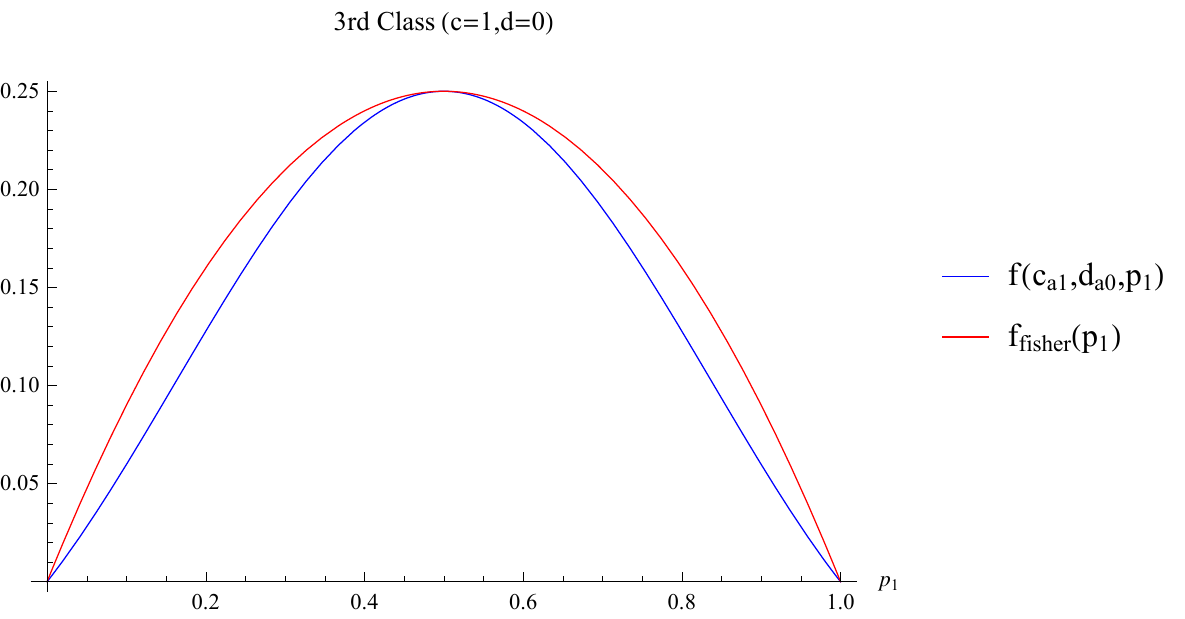}}
\caption{3rd class : (c,d) = (1,0)}
	\label{fig:2D class 3}
\end{figure}

\begin{figure}[htp]
	\centering
\scalebox{0.7}{\includegraphics[angle=0]{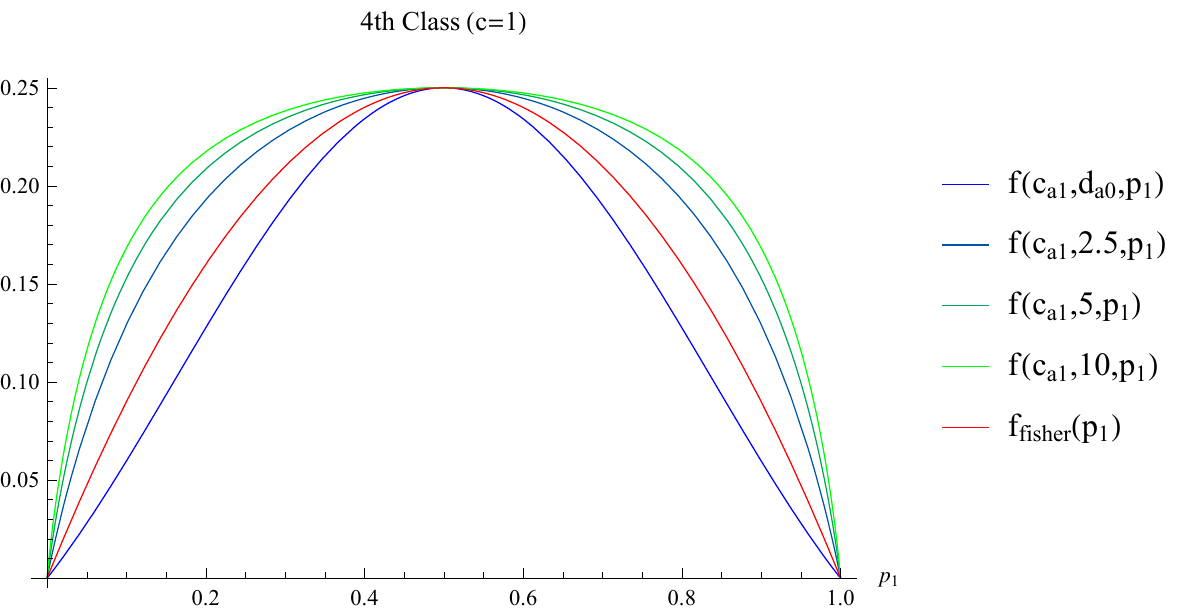}}
\caption{4th class : (c,d ) = (1,0) , (1,2.5) , (1,5) , (1,10)}
	\label{fig:2D class 4}
\end{figure}

\begin{figure}[htp]
	\centering
\scalebox{0.7}{\includegraphics[angle=0]{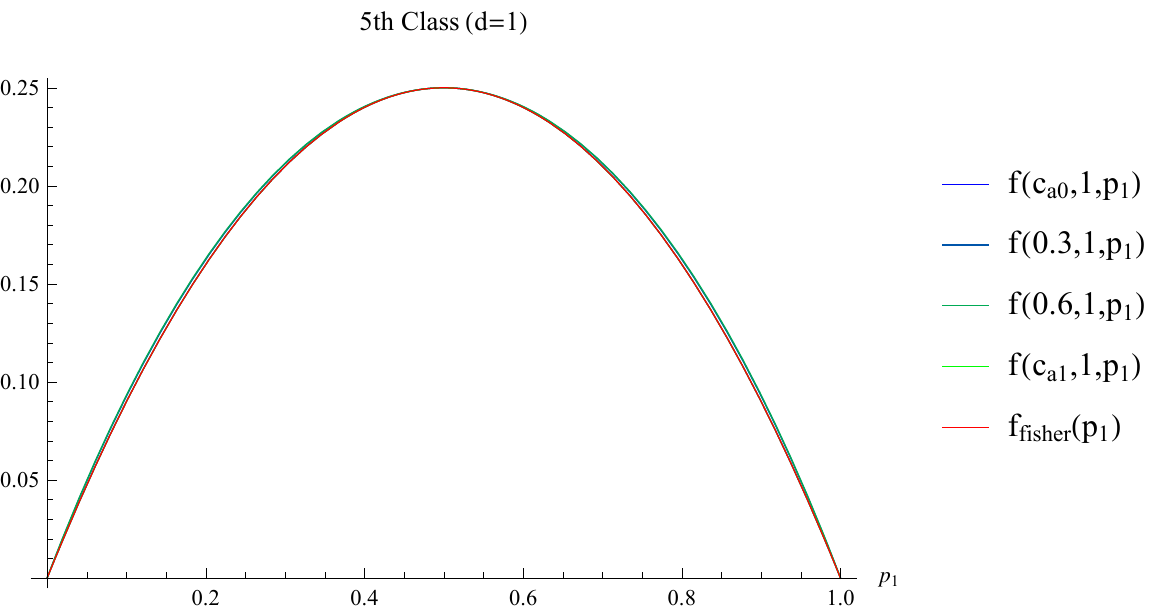}}
\caption{5th class : (c,d) = (0,1) , (0.3,1) , (0.6,1) , (1,1)}
	\label{fig:2D class 5}
\end{figure}

\begin{figure}[htp]
	\centering
\scalebox{0.7}{\includegraphics[angle=0]{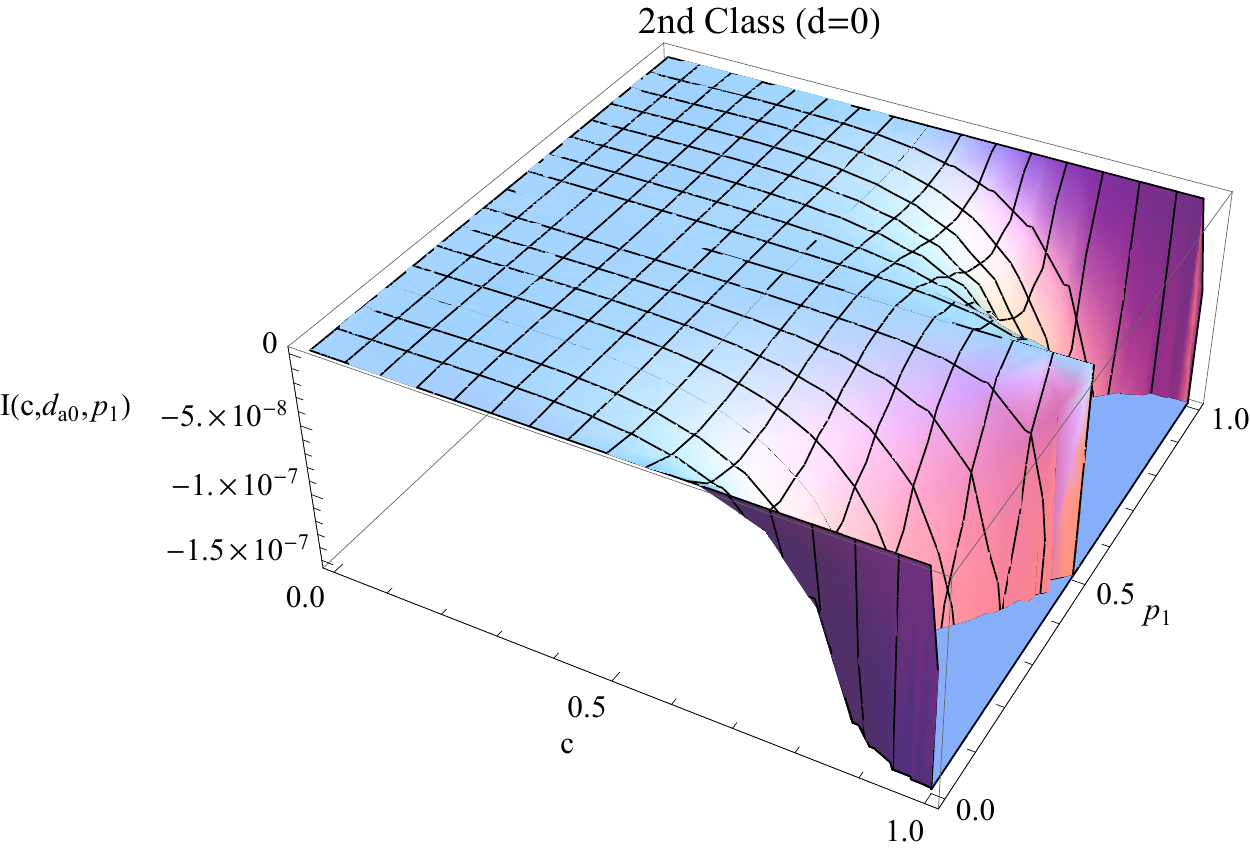}}
\caption{2nd class : I(c,d,$p_{1}$) = f(c,d,$p_{1}$) - $f_{Fisher}(p_{1})$ , d = 0 , $0<c<1$ }
	\label{fig:3D class2 diff}
\end{figure}

\begin{figure}[htp]
	\centering
\scalebox{0.7}{\includegraphics[angle=0]{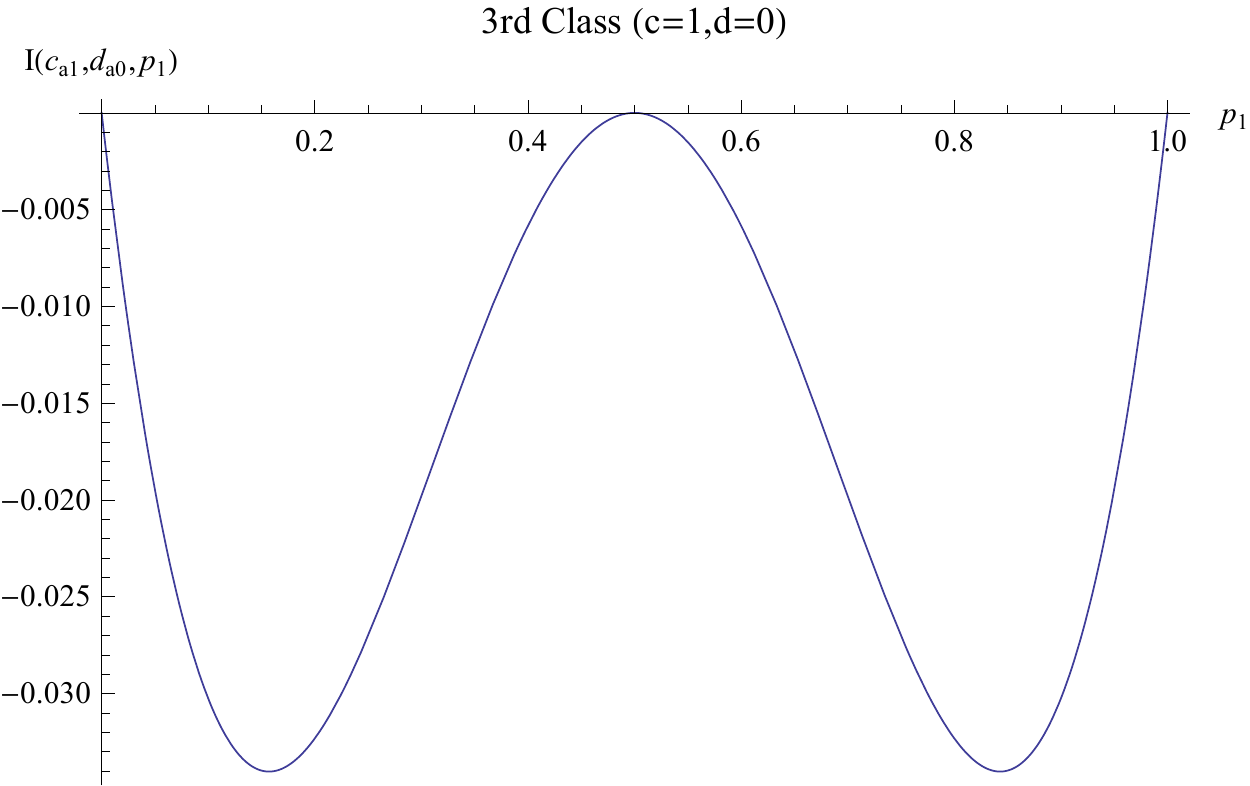}}
\caption{3rd class : I(c,d,$p_{1}$) = f(c,d,$p_{1}$) - $f_{Fisher}(p_{1})$ , c = 1, d = 0}
	\label{fig:2D class3 diff}
\end{figure}
\begin{figure}[htp]
	\centering
\scalebox{0.7}{\includegraphics[angle=0]{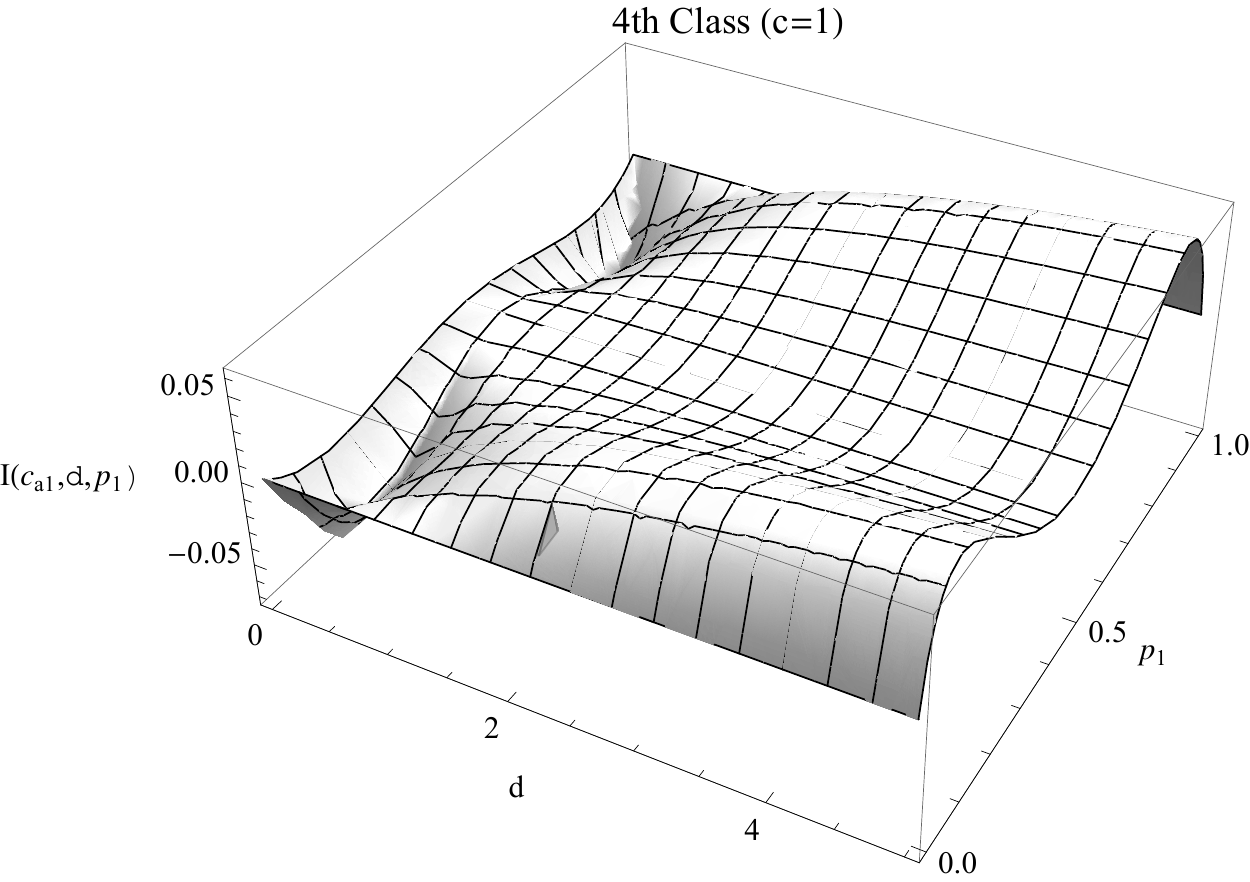}}
\caption{4th class : I(c,d,$p_{1}$) = f(c,d,$p_{1}$) - $f_{Fisher}(p_{1})$ , c = 1, $d>0$}
	\label{fig:3D class4 diff}
\end{figure}

\begin{figure}[htp]
	\centering
\scalebox{0.7}{\includegraphics[angle=0]{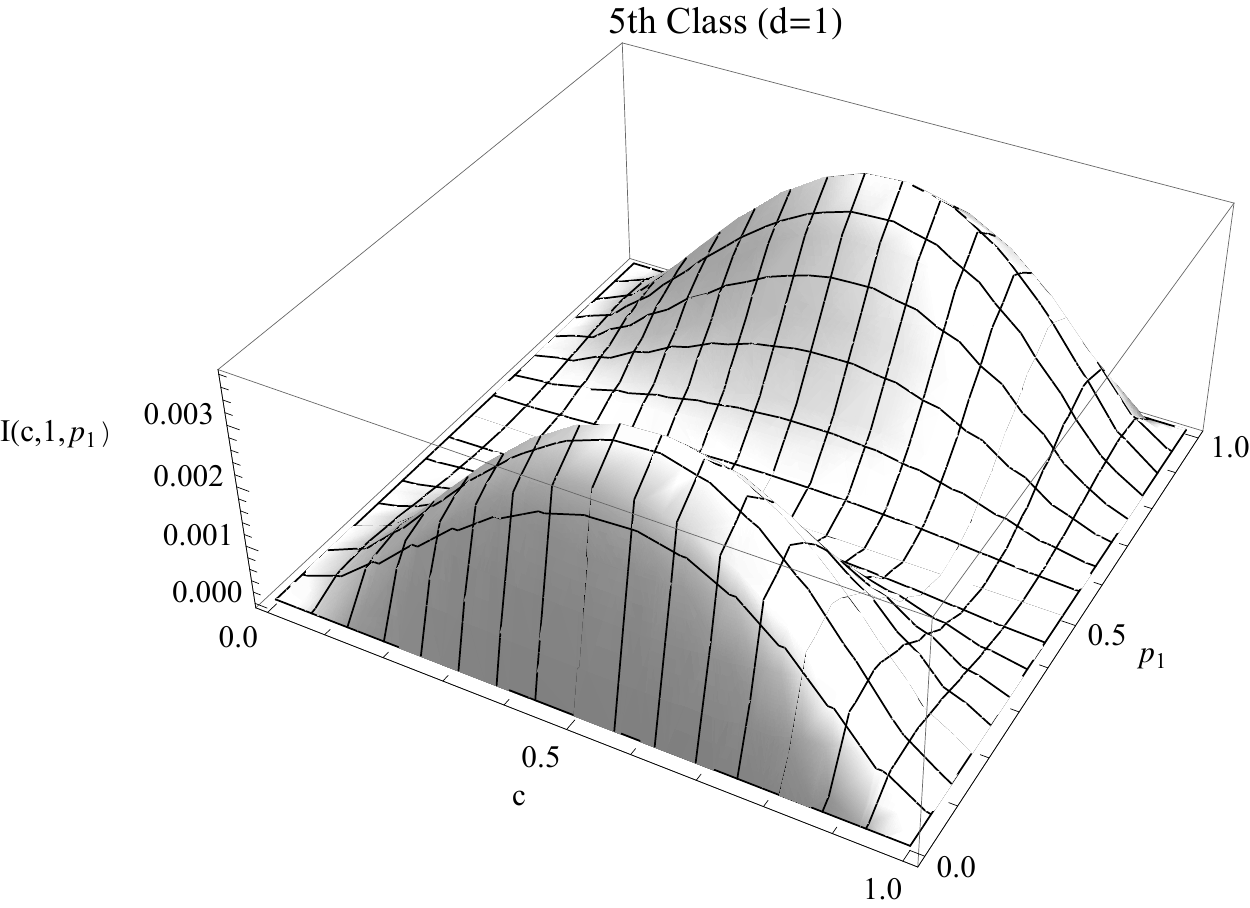}}
\caption{5th class : I(c,d,$p_{1}$) = f(c,d,$p_{1}$) - $f_{Fisher}(p_{1})$ , $0<c<1$  , d = 1}
	\label{fig:3D class5 diff}
\end{figure}

\begin{table}
  \begin{center}
    \begin{tabular}{|l|r|r|r|r|r|}
\hline
class &  c        &  d      &  $I_{mean}$ \\  \hline
1st   &  1        &  1      &   0         \\  \hline
2nd   &  $0<c<1$  &  0      &   0.0000    \\  \hline
3rd   &  1        &  0      &  -0.0184    \\  \hline
4th   &  1        &  $d>0$  &  0.0465    \\  \hline
5th   &  $0<c<1$  &  1      &   0.0012    \\  \hline
    \end{tabular}
    \caption{$I_{mean}$ is the average value of the difference between the Cramer-Rao bound for the Fisher metric and that of the (c,d)-metric. The average is over the values of the parameters that are not fixed but belong to given intervals. The table is meant to give a more quantitative differentiation between the 5 classes. In the 4th class the value of $I_{mean}$ is approximate}
  \end{center}
\end{table}


\subsection{Scalar Curvature Characterization}
\indent
For the discrete distribution the connection coefficients computed from the (c,d)-divergence turn out to be identically zero. This means that we have, as expected,  dually flat manifolds for all values of the parameters c,d. But the Levi-Civita connection and the resulting curvatures are not in general zero. We computed this curvature for the manifold point where all the $p_{i}'s$ are equal and plotted the ratio of this with the corresponding curvature given by the Fisher metric. Our diagrams show a clear dependence on the values of the c,d parameters. In the Hanel and Thurner papers it is stressed that their classification concerns the corresponding thermodynamic limits. This would mean that our information manifolds would be of infinite dimension. This is a very interesting challenge and we are working on this questions. Here we are very modest. In the Cramer-Rao analysis we investigate the behavior of very small systems, something provocatively far from the Hanel and Thurner assumption. But, as it turned out even in this hopeless case of small systems we do observe non trivial dependence on the c,d parameters. In the case of scalar curvature we try to be closer to their thesis. We have plotted the scalar curvature ratios for increasing dimensionality. We observe a form of saturation. Of course this is only a  hope that the number 12 say, is close to the infinity of the thermodynamic limit. Nevertheless the plots do differentiate between various classes in a clear way. We hope to obtain some asymptotic analytic estimates.

\begin{figure}[htp]
	\centering
\scalebox{0.7}{\includegraphics[angle=0]{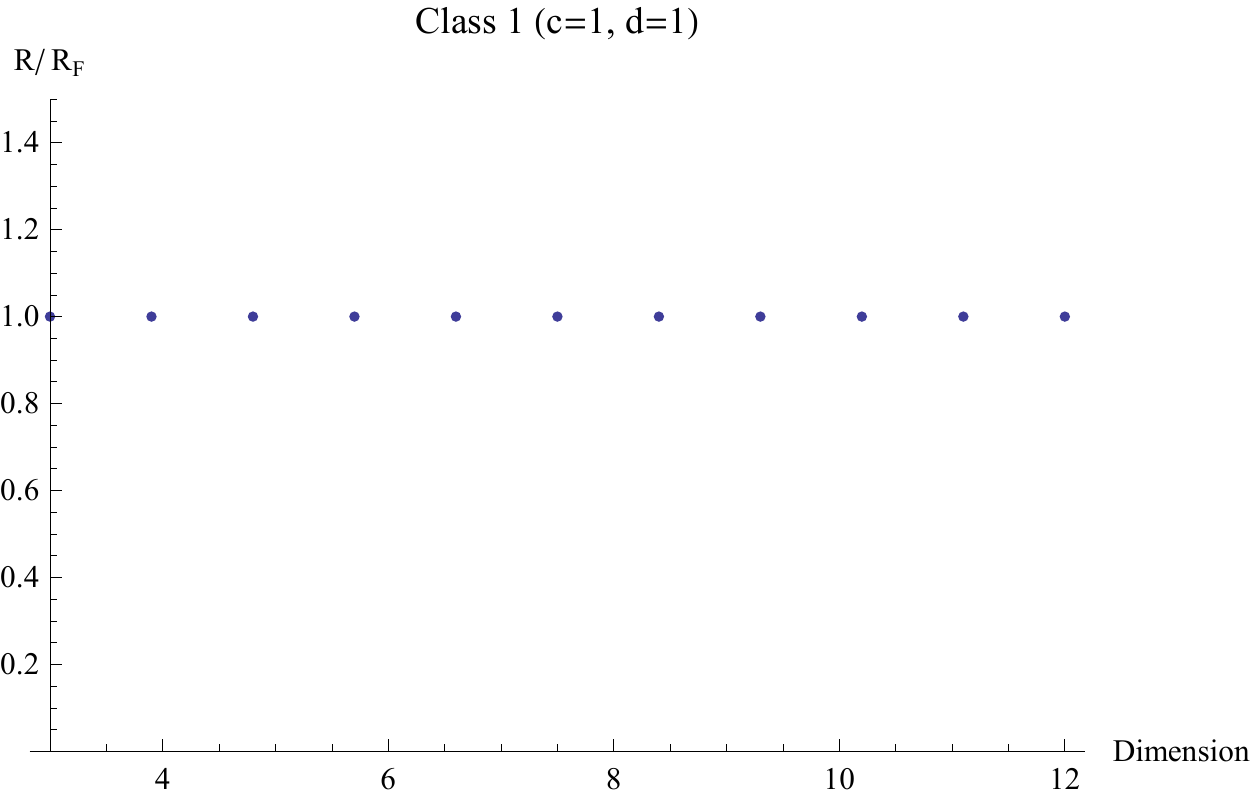}}
\caption{1st class : Ratio of (c,d) scalar curvature over Fisher scalar curvature }
	\label{fig:class 1 scalar}
\end{figure}

\begin{figure}[htp]
	\centering
\scalebox{0.7}{\includegraphics[angle=0]{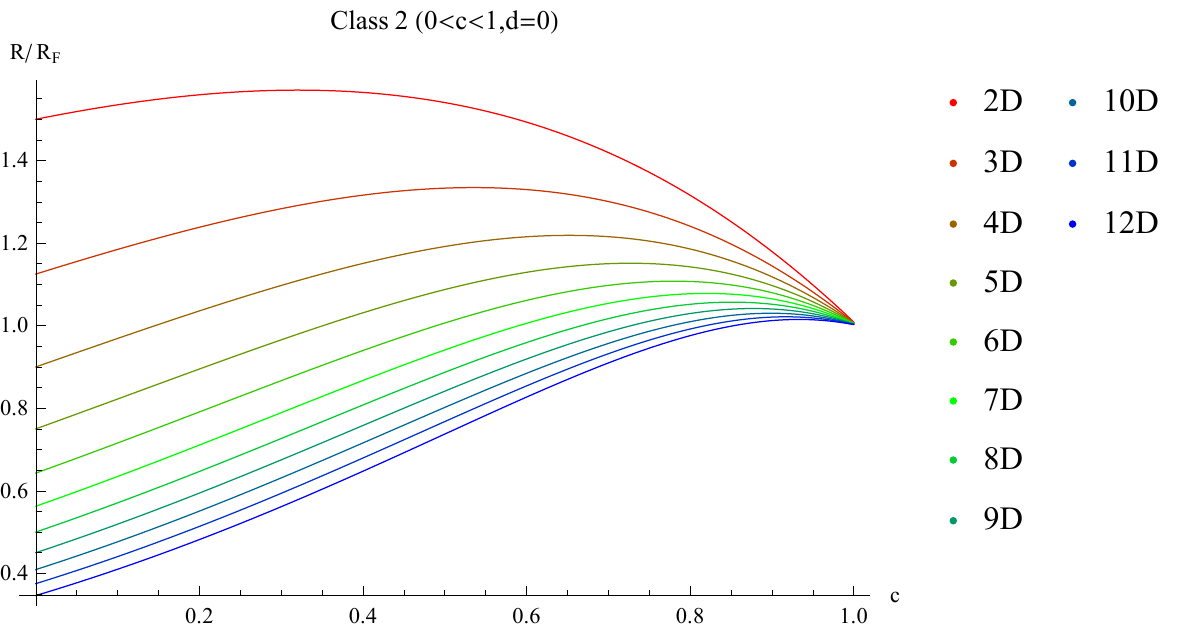}}
\caption{2nd class : Ratio of (c,d) scalar curvature over Fisher scalar curvature}
	\label{fig:class 2 scalar}
\end{figure}

\begin{figure}[htp]
	\centering
\scalebox{0.7}{\includegraphics[angle=0]{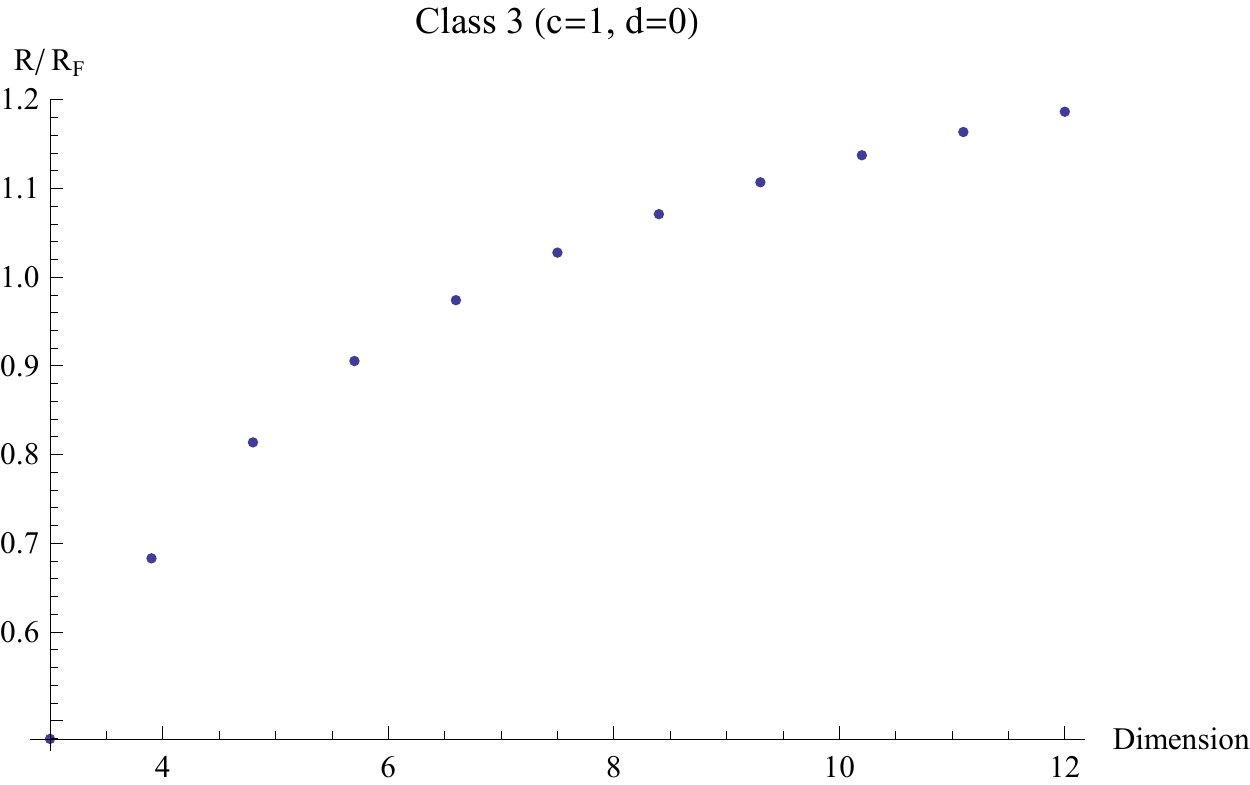}}
\caption{3rd class : Ratio of (c,d) scalar curvature over Fisher scalar curvature}
	\label{fig:class 3 scalar}
\end{figure}

\begin{figure}[htp]
	\centering
\scalebox{0.7}{\includegraphics[angle=0]{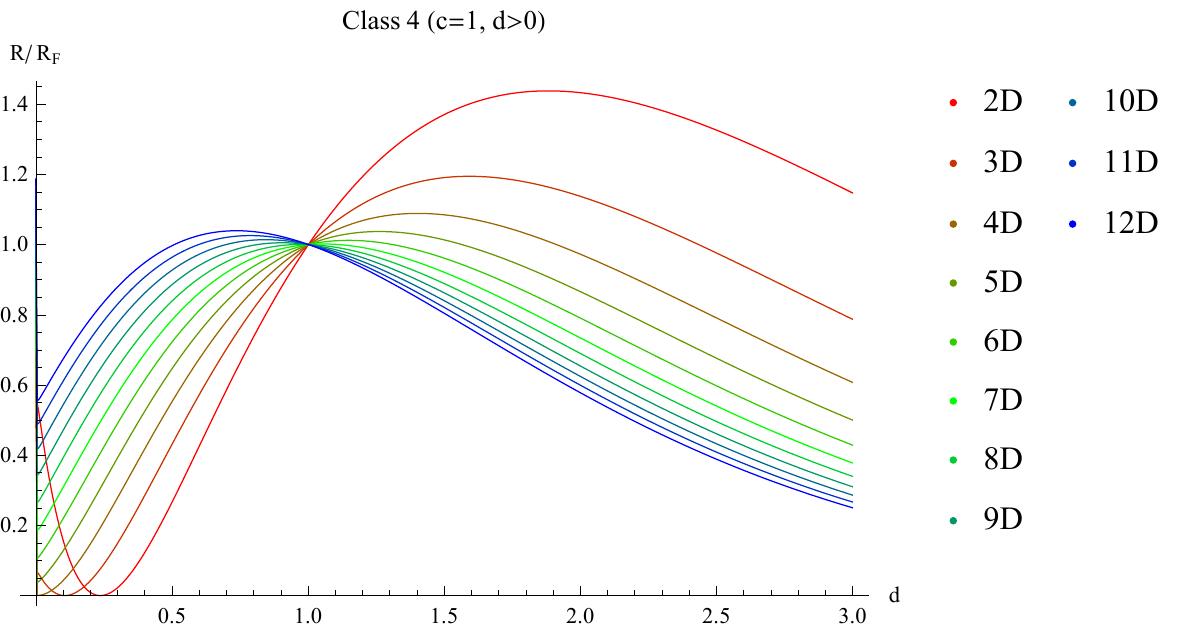}}
\caption{4th class : Ratio of (c,d) scalar curvature over Fisher scalar curvature}
	\label{fig:class 4 scalar}
\end{figure}

\begin{figure}[htp]
	\centering
\scalebox{0.7}{\includegraphics[angle=0]{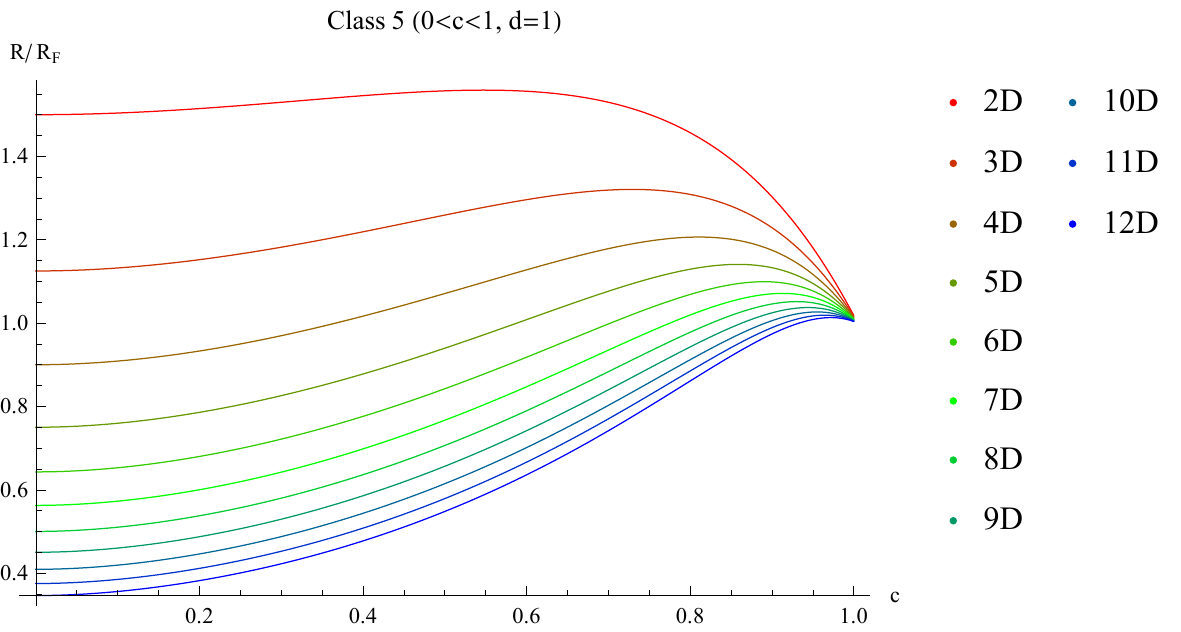}}
\caption{5th class : Ratio of (c,d) scalar curvature over Fisher scalar curvature}
	\label{fig:class 5 scalar}
\end{figure}

\section{Discussion}
\indent
We posed the question whether the generalized (c,d)-entropy of Hanel and Thurner \cite{hanel1,hanel2,hanel3}, and the corresponding generalized exponetial may give information manifolds that show clear dependence on these parameters.We proved four Theorems following the analysis of Amari \cite{amari1,amari2} and Naudts \cite{naud1,naud2}. We analyzed the dependence of the Cramer-Rao bounds and of the Levi-Civita scalar curvatures. For the bounds, to obtain a visual classification, we studied the extreme case of one dimensional manifold which, of course, is 'infinitely' far  from the thermodynamic limit of Hanel and Thurner, but still offers a first impression of the classification. Indeed the 2d diagrams and the associated Table show this (more or less). The 3d diagrams are more informative. The classification is more obvious.
 Further on increasing the dimensionality of the manifolds we observe a tendency which is characteristic for the different classes  in the diagrams of the scalar curvature. Here the differentiation between the classes is more revealing.
 Concerning the interpretation of these information theoretic quantities and their relevance for the classification of complex systems we have to appeal to the estimation theory. It might mean that various classes of complex systems (a la Hanel and Thurner) present differences in the estimation procedures of their parameters. We understand that this is a very strong statement given the fact that we study here very small systems, but we believe that we see a trend. We need further analysis and some rigorous inequalities related to the thermodynamic limit. The results for the curvature seem to be more safe with respect to the dimensionality, but the statistical use of the curvature is not very clear to us. We are analyzing this issue presently. 
\section{Appendix}
\indent
The proofs of the theorems are straightforward algebraic manipulations of properly defined functions. We repeat the definitions of the main text for convenience. We present only the necessary details.
\newline
\indent
\begin{bfseries}Theorem 1 \end{bfseries}
\indent
\begin{proof}
For a discrete distribution we have $p(x) = \sum_{i=0}^{n}p_{i}\delta_{i}(x)$. Then for an arbitrary function f(x) we get  
$f(p(x)) = f(\sum_{i=0}^{n}p_{i}\delta_{i}(x)) = \sum_{i=0}^{n}f(p_{i})\delta_{i}(x)$. Thus for $f(x) = \Lambda_{c,d,r}(p(x))$ 

\begin{eqnarray*}
\Lambda_{c,d,r}(p(x)) &=& r-r\Biggl[p_{0}^{c-1}\Biggl(1-\frac{1-(1-c)r}{dr}ln p_{0}\Biggr)^{d}\Biggr]\delta_{0}(x) \\
                      &-& r\sum_{i=1}^{n}\Biggl[p_{i}^{c-1}\Biggl(1-\frac{1-(1-c)r}{dr}ln p_{i}\Biggr)^{d}\Biggr]\delta_{i}(x)
\end{eqnarray*}

Using $\delta_{0}(x) = 1-\sum_{i=1}^{n}\delta_{i}(x)$ we have

\begin{eqnarray*}
\Lambda_{c,d,r}(p(x)) &=& r  -r\Biggl[p_{0}^{c-1}\Biggl(1-\frac{1-(1-c)r}{dr}ln p_{0}\Biggr)^{d}\Biggr] \\           
                      &+& \sum_{i=1}^{n}\Biggl\{r\Biggl[p_{0}^{c-1}\Biggl(1-\frac{1-(1-c)r}{dr}ln p_{0}\Biggr)^{d}\Biggr]  \\
                      &-&r\Biggl[p_{i}^{c-1}\Biggl(1-\frac{1-(1-c)r}{dr}ln p_{i}\Biggr)^{d}\Biggr]\Biggr\}\delta_{i}(x) \\
										  &\equiv& \sum_{i=1}^{n}x_{i}\theta^{i} - \psi(\theta)
\end{eqnarray*}
\end{proof}

\begin{bfseries}Theorem 2 \end{bfseries}
\indent
\begin{proof}
We define the function
\begin{equation}
\Delta(x) = \frac{1}{r(1-c)}\mathcal{E}_{c,d,r}(x)\frac{W(B(1-x/r)^{1/d})}{1+W(B(1-x/r)^{1/d})}\Biggl(1-\frac{x}{r}\Biggr)^{-1}
\end{equation}
Differentiating w.r.t. $\theta$ we get
\begin{equation}
\frac{\partial\psi}{\partial\theta^{i}} \equiv \partial_{i}\psi = \frac{\int x_{i}\Delta(x_{j}\theta^{j}-\psi(\theta))dx}{\int\Delta(x_{j}\theta^{j}-\psi(\theta))dx}
\end{equation}
and
\begin{equation}
\partial_{k}\partial_{i}\psi(\theta) = \frac{\int \Delta^{'}(x_{j}\theta^{j}-\psi(\theta))(x_{k}-\partial_{k}\psi)(x_{i}-\partial_{i}\psi)dx}{\int \Delta(x_{j}\theta^{j}-\psi(\theta))dx}
\end{equation}
From the latter expression it follows that $\psi(\theta)$ is convex for the stated conditions on $\Delta$ and $\Delta^{'}$.
\end{proof}

\begin{bfseries}Theorem 3 \end{bfseries}
\indent
\begin{proof}
Defining the function
\begin{equation}
K(x) = x^{\frac{c-1}{d}}\Biggl( 1 - \frac{1-(1-c)r}{rd}logx \Biggr)
\end{equation}
so that $\Lambda_{c,d,r}(x) = r-rK^{d}(x)$ and the function
\begin{equation}
h(p) = \int \Delta(\Lambda(p(x,\theta)))dx \equiv \int \Delta(x_{i}\theta^{i} - \psi(\theta))de
\end{equation}
we get
\begin{equation}
h(p) = \sum_{i=0}^{n}p_{i}\frac{1}{r(1-c)}\frac{W(BK(p_{i}))}{1 + W(BK(p_{i}))}K^{-d}(p_{i})
\end{equation}
The canonical (c,d)-divergence is
\begin{equation}
D_{c,d,r}(p(x,\theta_{1}),p(x,\theta_{2})) = \psi(\theta_{2})-\psi(\theta_{1}) - \partial_{i}\psi(\theta_{1})(\theta_{2}^{i}-\theta_{1}^{i})
\end{equation}
For discrete distributions , namely $x_{i}= \delta_{i}(x)$ and $p(x,\theta_{1}) = (p_{1},...,p_{n})$ , $p(x,\theta_{2}) = (q_{1},...,q_{n})$
we have
\begin{equation}
\partial_{i}\psi = \frac{\Delta(\theta^{i}-\psi(\theta))}{h(p)}
\end{equation}
and the divergence becomes
\begin{eqnarray*}
D_{c,d,r}(p,q) &=& -\Lambda_{c,d,r}(q_{0}) +  \Lambda_{c,d,r}(p_{0}) - \sum_{i=1}^{n}\frac{\Delta(\theta_{1}^{i})-\psi(\theta_{1})}{h(p)}(\theta_{2}^{i} - \theta_{1}^{i})   \\        &=& -r + rq_{0}^{c-1}\Biggl(1 - \frac{1-(1-c)r}{dr}lnq_{0}\Biggr)^{d} \\
               &+&r - rp_{0}^{c-1}\Biggl(1 - \frac{1-(1-c)r}{dr}lnp_{0}\Biggr)^{d} \\
               &-& \sum_{i=1}^{n}\frac{\Delta(\theta_{1}^{i}-\psi(\theta_{1}))}{h(p)}(\theta_{2}^{i}-\theta_{1}^{i}) 
\end{eqnarray*}
Using Theorem 1 we have after some substitutions for discrete distributions
\begin{eqnarray*}
D_{c,d,r}(p,q) &=& rK^{d}(q_{0}) - rK^{d}(p_{0}) \\
               &-& \frac{1}{r(1-c)}\frac{1}{h(p)}\sum_{i=1}^{n}p_{i}\frac{W(BK(p_{i}))}{1 + W(BK(p_{i}))}K^{-d}\\
							 &\times& \biggl\{ rK^{d}(q_{0}) - rK^{d}(q_{i}) - rK^{d}(p_{0}) + rK^{d}(p_{i}) \biggr\}
\end{eqnarray*}
We obtain the required result by extending the sum to $i=0$.
\end{proof}

\begin{bfseries}Theorem 4 \end{bfseries}
\indent
\begin{proof}
The metric is derived from the expansion of the divergence
\begin{equation}
g_{ij}(p) = \frac{\partial^{2}}{\partial q_{i}\partial q_{j}}D_{c,d,r}(p,q)|_{q=p}
\end{equation}
The result is obtained by a series of simple manipulations of  the expressions defined in the previous theorems.
\end{proof}

\footnote{One of the authors (F.O.) wishes to thank the Greek State Scholarships Foundation for a research scholarship.}


\begin{thebibliography}{99}
\bibitem{hanel1} Hanel R. and Thurner S. 2011  {\it EPL} {\bf 93} 20006.
\bibitem{hanel2} Hanel R. and Thurner S. 2013 , arXive : 1310.5959v1 [cond-mat.stat-mech]
\bibitem{hanel3} Hanel R. and Thurner S. 2013 {\it Is there a world behind Shannon? Entropies for complex systems} in {\it Interdisciplinary Symposium on Complex Systems} Emergence, Complexity and Computation , A. Sanyaei et al (eds).
\bibitem{cencov} Chencov N N 1982 {\it Statistical Decision Rules} (Rhode Island USA : AMS)
\bibitem{amari1} Amari Shun-ichi and Nagaoka H 2000 {\it Methods of Information Geometry}  (Oxford University Press : AMS)
\bibitem{nicol1} Nicolis G. and Nicolis C. 2007 {\it Foundations of Complex Systems}, World Scientific, Singapore.
\bibitem{amari2} Amari Shun-ichi and Atsumi Ohara 2011 {\it Entropy} {\bf 13} 1170-1185.
\bibitem{lady} Ladyman J., Lambert J. Wiesner K. 2012 {\it What is a Complex System?} , preprint.
\bibitem{mag} Magee C.L. , de Weck O.L. , 2004 {\it Complex System Classification} in Fourteen Annual International Symposium of the International Council on Systems Engineering (INCOSE).
\bibitem{zad} Zadeh L.A. , 1973{\ IEEE Transactions on Systems, Man, and Cybernetics},  {\bf SMC-3} , 28.
\bibitem{mere} Merelli E. , Rucco M. , Sloot P. , Tesei L. 2015, {\it Entropy} {\bf 17} 6872-6892.  
\bibitem{monc} Moncion T. , Amar P. , Hutzler G. 2010 , {\it Journal of Biological Physics and Chemistry}, {\bf 10}. 
\bibitem{gaud} Gaudiano M.E. 2015, {\it Physica A: Statistical Mechanics and its Applications}, {\bf 440}, 185-199.
\bibitem{koore} Koorehdavoudi H., Bogdan P. 2016, {\it A Statistical Physics Characterization of Complex Systems Dynamics: Quantifying Complexity from Spatio-Temporal Interactions} , Scientific Reports {\bf 6}, Article number 27602.
\bibitem{rotten} Rottenberg S., Leriche S. , Taconet C., Lecocq C. , Desprats T. ,2014,{\it MuSCa: A Multiscale Characterization Framework for Complex Distributed Systems } in Proceedings of the 2014 Federal Conference on Complex Science and Information Systems pp 1657-1665.
\bibitem{tsall1} Tsallis C., Mendes R.S. , Plastino A.R. 1998, {\it Physics A} {\bf 261} , 534-554.
\bibitem{tsall2} Tsallis C. 2009, {\it Introduction to non-extensive statistical mechanics}, Springer.
\bibitem{tsall3} Tsallis C. 2004, {\it Non-extensive statistical mechanics: construction and physicsl interpretation}, in M. Gell-Mann, C. Tsallis (eds) Non-extensive Entropy pp. 1-53, Oxford University Press, Oxford 2004.
\bibitem{naud1} Naudts J. 2004, {\it Estimators, escort probabilities, and $\phi$-exponential functions in statistical physics}, arXive: math-ph/0402005v1.
\bibitem{naud2} Naudts J. 2011, {\it Generalized Thermostatistics} , Springer. 
\bibitem{wada}  Wada T. Scarfone A.M. , 2015, {\it Entropy} {\bf 17} 1204-1217.
\bibitem{mats} Matsuzone H. , Wada T. , 2015, {\it Entropy} {\bf 17} 5729-5751.
\bibitem{hars} Harsha K.V. , Subrahamanian Mooshath K.S. ,2015, {\it Physics A} {\bf 433} , 135-147.
\bibitem{beck} Beck C. ,2003, {\it Superstatistics, escort distributions, and applications}, arXiv: cond-mat/0312134v1.
\bibitem{abe} Abe S. , 2003, {\it Geometry of escort distributions}, \\arXive: cond-mat/0305231.
\bibitem{penn} Pennini F. , Plastino A. ,2004,  {\it Escort-Husimi distributions, Fisher information and non-extensivity }, arXiv: cond-mat/0402467v2.
\bibitem{masi} Masi M. , 2006, {\it Generalized information-entropy measures and Fisher information}, arXiv: cond-mat//0611300v2.
\bibitem{oik} Oikonomou T. Baris Bagci G. , 2009, {\it The maximization of Tsallis entropy with complete deformed functions and the problem of constrains}, arXiv: cond-mat.stat-mech//0907.4059v1.
\bibitem{late} Latella I. , Perez-Madrid A., Campa A., Casetti L., Ruffo S., 2015,  arXiv: cond-mat.stat-mech//1505.03767v1.
\bibitem{ber} Jean-Francois Bercher, 2010 , {\it On escort distributions, q-gaussians and Fisher Information}, , in 30th International Workshop on Bayesian Inference and Maximum Entropy Methods in Science and Engineering, Jul 2011, Chamonix, France.
\bibitem{ay} Ay Nihat, 2011, {\it A Geometric Approach to Complexity}, SFI WORKING PAPER 2011-08-039.
\end{thebibliography}
\end{document}